\documentclass{nature}
\usepackage{graphicx}
\usepackage{dcolumn}
\usepackage{bm}
\usepackage{amsmath}
\usepackage{mathtools}

\usepackage{newfloat}
\DeclareFloatingEnvironment[name={Extended Data Figure}]{exfigure}

\title{Three-dimensional imaging of dislocation propagation during crystal growth and dissolution}

\author{Jesse N. Clark$^{1,\#,\dagger}$, Johannes Ihli$^{2,\dagger}$, Anna S. Schenk$^{2}$, Yi-Yeoun Kim$^{2}$, Alexander N. Kulak$^{2}$, James M. Campbell$^{3}$, Gareth Nisbet$^{4}$, Fiona C. Meldrum$^{2}$, \& Ian K. Robinson$^{1,5}$}

\begin{document}

\maketitle

\begin{affiliations}
 \item London Centre for Nanotechnology, University College, London WC1E 6BT, UK
\item  School of Chemistry, University of Leeds, Leeds LS2 9JT, UK.
\item School of Physics and Astronomy, University of Leeds, Leeds, LS2 9JT, UK
\item Diamond Light Source, Harwell Science and Innovation Campus, Didcot, Oxon OX11 0DE, UK
 \item Research Complex at Harwell, Didcot, Oxfordshire OX11 0DE, UK \\
\# Present address: Stanford PULSE Institute, SLAC National Accelerator Laboratory, 2575 Sand Hill Road, Menlo Park, California 94025, USA and Center for Free-Electron Laser Science (CFEL), Deutsches Elektronensynchrotron (DESY), Notkestrasse 85, 22607 Hamburg, Germany. \\
 $\dagger$ These authors contributed equally to this work \nonumber

 
\end{affiliations}

\begin{abstract}
Atomic level defects such as dislocations play key roles in determining the macroscopic properties of crystalline materials \cite{Stoneham:1985aa,Burton:1892aa}.  Their effects are important and wide-reaching, and range from increased chemical reactivity \cite{Lasaga:2001aa,De-Yoreo:2003aa}, to enhanced mechanical properties \cite{Ma:2006aa,Kunitake:2013aa}, to vastly increased rates of crystal growth.  Dislocations have therefore been widely studied using traditional techniques such as X-ray diffraction (XRD) and optical imaging.  More recently, advances in microscopy have allowed their direct visualization.  Atomic force microscopy (AFM) has enabled the 2D study of single dislocations \cite{Davis:2000aa} while transmission electron microscopy (TEM), which was initially limited to 2D projections of thin specimens, can now visualize strain fields in 3D with near atomic resolution \cite{Hytch:2003aa,Barnard:2006aa,Chen:2013aa}.  However, these techniques cannot offer in situ, 3D imaging of the formation or movement of dislocations during dynamic processes such as crystal growth and dissolution.  Here, we describe how Bragg Coherent Diffraction Imaging (BCDI) \cite{Pfeifer:2006aa,Robinson:2009aa} can be used to visualize in 3D the entire network of dislocations present within an individual crystal.  Using calcite (CaCO$_3$) single crystals, we also use BCDI to monitor the propagation of the dislocation network during repeated growth and dissolution cycles, and show how this is intimately linked to the growth and dissolution mechanisms.  These investigations demonstrate the potential of BCDI for studying the mechanisms underlying the response of crystalline materials to external stimuli.

\end{abstract}
Crystal growth and dissolution processes have been studied for over a century \cite{Weeks:2007aa}, due to their significance to fields such as geology, corrosion, catalysis and the synthesis of nano-structures, through a desire to understand the link between microscopic and macroscopic processes, and of course due to our innate fascination with such beautiful structures.  These investigations confirmed that dissolution and growth proceed by analogous mechanisms \cite{Dove:2007aa} and also identified that crystallographic defects, and in particular screw dislocations, are of fundamental importance to crystal growth and dissolution processes \cite{Frank:1949aa}.  This can be attributed to the fact that screw dislocations cause deformation (and therefore strain) of the adjacent crystal lattice, which in turn alters the activation barrier for growth and dissolution in the vicinity of the dislocation \cite{Brantley:2008aa}.  A full picture of crystal growth and dissolution mechanisms can therefore only be obtained by studying the relationship between the evolution of the network of dislocations within a crystal and its morphology during these processes.  
\\
\\
In this article, we demonstrate how BCDI \cite{Clark:2013aa,Cha:2013aa} can be used to study the role of dislocations in dictating the mechanism of growth and dissolution of calcite crystals.  Calcite was selected for study as it is one of the most-studied inorganic crystals, and the crystals readily grow to a few microns in size, with well-defined morphologies.  BCDI is an imaging technique that uses coherent X-rays to image the density (and morphology) of a crystal, and importantly, the strain within it \cite{Pfeifer:2006aa,Robinson:2009aa}.  Illumination of a crystal that is smaller than the coherence volume of the beam generates a coherent X-ray diffraction (CXD) pattern due to scattering from all parts of the crystal.  An image of the crystal morphology can then be generated from the CXD pattern, where the phase of the scattered wave is reconstructed using iterative phase retrieval algorithms \cite{Fienup:1982aa,Pfeifer:2006aa,Robinson:2009aa,Clark:2013aa,Cha:2013aa}.  The reconstructed density is complex valued, with the amplitude containing information about the electron density,  $\rho(\bm{r})$.  Phase shifts in the reconstructed complex density arise from strain (internal deformation) in the crystal lattice.  The phase is proportional to the vector displacement field, $\bm{u}(\bm{r})$ of the atoms from the ideal lattice points and the scattering vector $\bm{Q}$ via $\phi(\bm{r})= \bm{Q} \cdot \bm{u}(\bm{r})$ (see supplementary information for further details). For a single Bragg peak a single projection of $\bm{u}$ is obtained (projected displacement) and components of $\bm{u}$ perpendicular to $\bm{Q}$ will not be observed.   It is this sensitivity to deformations that makes BCDI ideal for studying defects within crystals.
\\
\\
Calcite crystals were precipitated by placing 100 $\mu$L droplets of a solution containing CaCl$_2$, urea and urease on hydroxyl terminated, self-assembled monolayers (SAMs) supported on gold thin films.  CaCO$_3$ precipitation then occurred on enzymatic hydrolysis of the urea to ammonium and carbonate \cite{Antipov:2003aa}.  This method was selected as it generated a high density of $\{$104$\}$ oriented calcite rhombohedra with average diameters of $\approx$ 1.25 $\mu$m \cite{Lee:2007aa}.  BCDI experiments were then carried out at beamline I16 at Diamond Light Source.  Individual calcite crystals below 2 $\mu$m in size were illuminated with monochromatic, 8 keV X-rays, and diffraction was recorded at the $\{$104$\}$ Bragg peak.  3D diffraction data sets (Supplementary Information Fig. 1) were obtained by rocking an isolated calcite crystal through its Bragg peak, and the same crystal was monitored while undergoing cycles of growth and dissolution.  Dissolution was achieved by depositing dilute acetic acid solution on the crystal, while growth was achieved by adding a drop of calcium bicarbonate solution.  Alignment of the crystal was maintained throughout, as the X-rays were nominally unfocussed and defined by slits with a square opening of 200 $\mu$m placed 0.3 m before the sample.  This large beam size relative to crystal size also ensured that the sample was coherently illuminated \cite{Clark:2012aa}.
\\
\\
Figure 1 shows 3D images of the crystals as iso-surface renderings of the reconstructed amplitudes (electron density, see supplementary information for details) (a) and phase (projected lattice displacement) (b) of the initial crystal (i), after growth (ii) and after two consecutive dissolution steps (iii and iv).  The initial crystal (i) is the expected rhombohedron, where this is consistent with scanning electron micrographs (Supplementary Information Fig. 3).  Growth of this crystal (i to ii) leads to an increase in size and a smoothing of the faces exposed to the bulk solution, while the face in contact with the SAM remains unchanged.  Interestingly, two of the faces directed into the solution (indicated with blue arrows) grow more rapidly than the other three.  This is immediately indicative of a non-uniform distribution of defects, as all faces of an entirely perfect rhombohedron would be expected to grow at the same rate.  Images of the corresponding projected displacements are shown in Figure 1b, where this is mapped onto an iso-surface with red and blue representing lattice contraction or expansion respectively by half a lattice spacing.  Comparison of the projected displacements before (i) and after (ii) crystal growth shows that these do not grow significantly with the crystal but remain maximal at the edges.  This is indicative of the presence of active growth fronts \cite{Paquette:1995aa}.
\\
\\
The dissolution steps (Fig. 1, ii to iii and iii to iv) show that the crystal faces retreat along all directions, but that this is more pronounced at certain sites (indicated by red arrows).  Initial signs of changes in the crystal shape and the onset of etch-pit formation are visible after the first stage of dissolution (iii), which leads to an increase in the specific surface area and roughness of the crystal.  The etch pits are also associated with higher levels of deformation/strain.  That relatively little change occurs in the crystal face adjacent to the substrate is consistent with the intimate contact of the SAM with this nucleation face.  The second dissolution step (Fig. 1 (iii-iv) and Fig. 5 (iii-iv)) results in a significant change in the crystal morphology, and the production of a porous isometric form \cite{Snyder:2007aa} that is quite distinct from the original shape and which can be attributed to the removal of defect outcrops at the crystal surface and the coinciding etch pits formed \cite{MacInnis:1992aa,MacInnis:1993aa}.  Looking in turn at the lattice deformation, it is evident that strain present at the crystal surface reduces with the increased dissolution.  This indicates that the least stable (more strained) regions dissolve first, leaving behind a more stable core (Supplementary Movies 1-4).  
\\
\\
The projected displacement images also reveal a further intriguing feature, which is indicated by the grey arrows in Fig. 1b, and which is present throughout the growth and dissolution of the crystal.  This region possesses both a hollow core and a spiral phase, where this combination of features is characteristic of screw dislocations; a number are highlighted in Fig. 2.  Dislocations are characterized by a Burgers vector that measures the topological shift of the crystal along the dislocation line, where this is usually a lattice vector of the crystal \cite{Hirth1968}.  Whenever there is a component of the Burgers vector parallel to the dislocation line, it has a screw dislocation character causing the lattice to spiral around the dislocation; in this way crystal growth (and dissolution) can be facilitated.  Confirmation that this feature indeed corresponds to a screw dislocation was obtained by recording the polar angle dependence of the displacement associated with the core (indicated by the circle in Figure 2a) over the growth/ dissolution cycle of the crystal (Figure 3a).  An approximately linear relationship was observed and is consistent with what is predicted by linear elasticity theory \cite{Hirth1968} (see supplementary information).  To further confirm the nature of the identified dislocation, a comparison is provided between a simulated screw dislocation and simulated screw after BCDI processing.  The model screw dislocation used in the simulation is shown at atomic resolution in Figure 4a, with the resulting displacement of atoms from their ideal lattice positions given in Figure 4c.  The BCDI experiment processed simulated screw dislocation is shown in Figures 4b and 4d (see supplementary information for details).  Figures 4e and 4f provide a comparison of BCDI reconstructed displacement and simulated displacement, viewed along the dislocation line, and clearly show that the low density core and spiral displacement are well-preserved after BCDI processing.
\\
\\
Further examination of regions that showed a spiral deformation and low-amplitude core enabled many additional dislocations to be identified within the imaged crystal (see supplementary information).  These have been rendered and are shown in Figure 5 and Supplementary Movies 5-8.  The initial crystal (i) possesses several dislocations which are located relatively close to the crystal surface (down to 200 nm) and are found predominantly parallel to $\{$104$\}$ planes.  These dislocations are mainly associated with the faces exposed to the bulk solution, and are visible throughout a single crystal growth and dissolution cycle (steps i-iii), during which they (ii) increase and (iii) decrease in length.  The reproducibility of locating the same dislocation across independently reconstructed data sets provides further credence to the recovered images.  
\\
\\
Dislocations are often found to occur in loops or pairs with opposite Burgers vector, as this reduces the lattice potential energy and resulting long-range strain associated with these features.  They also are stabilized near crystal surfaces, as found in this work,  such that  dislocation motion can transport material into and out of crystals from the growth solution \cite{PurjaPun2009}.  As intuitively expected, the most rapidly growing crystal faces were observed to have the highest number of surface dislocations.  The distribution of these defects within the crystal therefore plays a part in determining the morphology of the product crystal.  Considering the dissolution process, the locations of the dislocations coincide with the position of the etch pits that appear at the crystal surface during dissolution.  This suggests that the developing pits follow the cores of the dislocations, where this effect can be attributed to a reduced activation barrier to dissolution due to stored strain energy in the defects14.  New dislocations could be identified during the dissolution (iii to iv), and by the final stage (iv) many of the original dislocations had been annihilated.  The loss of faceting and dislocations near the surface support the hypothesis that the least stable regions dissolve first, leaving behind a more stable core.
\\
\\
To determine the overall effect that the growth/ dissolution had on the strain and deformation of the crystal, the gradient of the displacement was calculated and its magnitude was plotted as a function of fractional size (Figure 3b and c).  The magnitude of the gradient was calculated over successively larger shells and averaged over all directions and is plotted as a $\%$ of deformation (strain) relative to the lattice constant for the {104} reflection (see supplementary information for details) for the directions parallel and perpendicular to the scattering vector (which is approximately parallel with y).  What is evident (Fig. 3b and c) is that the initial crystallite experiences an increasing strain from the center, outwards.  This can be compared to a crystal formed after the first growth stage (ii), which shows an overall flatter initial strain, and then increases more rapidly above a fractional size $\approx$ 0.6.  The smoother transition and lower strain for the direction parallel suggest some relaxation of the crystal after the initial growth and can also be seen in the slices in Fig. 2.  These data are therefore consistent with the assertion that the effect of the surface only penetrates to a finite distance into a crystal.  With the increased roughness and etch-pit formation that occurs in the first dissolution step (iii), the overall strain is higher, demonstrating that these surface features affect the entire crystal, leading to an increase in overall deformation.  After the final dissolution (iv) when the majority of surface dislocations have been removed, the strain becomes almost flat.  This indicates that surface effects are less pronounced, despite the increased relative surface area.
\\
\\
With its ability to simultaneously generate 3D images of the strain within a crystal and the gross crystal morphology, BCDI provides an extremely powerful way of visualizing the network of dislocations present within an individual crystal. Importantly, BCDI can be performed in the absence of any sample preparation, which means that it can be used to elucidate the effects of dislocation networks on dynamic crystal behavior. Using the example of crystal growth and dissolution, we have located dislocations within calcite single crystals, and then demonstrated that their distribution within the crystal dictates rapidly-growing directions. Conversely, preferential dissolution and etch pit formation was also observed within the vicinity of the dislocations, which provides evidence that the stored energy within the dislocations affects the crystal response. The ability to view these fundamental processes using BCDI represents an important step forward in elucidating the nanoscale mechanisms underlying crystallization processes. Importantly, BCDI also opens the door to 3D visualization of the role of dislocations in the response of a crystal to a huge range of experimental conditions, such as temperature changes or mechanical force.

\section*{References}

\begin{addendum}
\item This work was supported by FP7 advanced grant from the European Research Council (J.N.C. and I.K.R.) and an Engineering and Physical Sciences Research Council Leadership Fellowship (F.C.M. and J.I.).  It was also funded through an EPSRC Programme Grant (A.S.S. and F.C.M., EP/I001514/1) which funds the Materials in Biology (MIB) consortium, and EPSRC grants EP/J018589/1 (YYK) and EP/K006304/1 (ANK).  We thank Diamond Light Source for access to Beamline I-16 (MT 8187, MT 7654 and MT 7277) that contributed to the results presented here.
\item[Author Contributions] J.N.C. and J.I. designed the project; J.I. prepared samples; J.N.C., J.I., J.M.C., A.S.S., Y.Y.K., J.M.C., G.N. and I.K.R performed the experiments; J.N.C. performed image reconstructions; J.N.C. $\&$ I.K.R. analyzed the Data, J.N.C., J.I., F.C.M. and I.K.R. wrote the paper. All the authors read and commented on the manuscript. Correspondence and requests for materials should be addressed to jesclark@stanford.edu or F.Meldrum@leeds.ac.uk.
\item[Financial Interest]
The authors declare no competing financial interest 

\end{addendum}

\clearpage

\begin{figure}
 	\centering
	\includegraphics[width=.96  \textwidth]{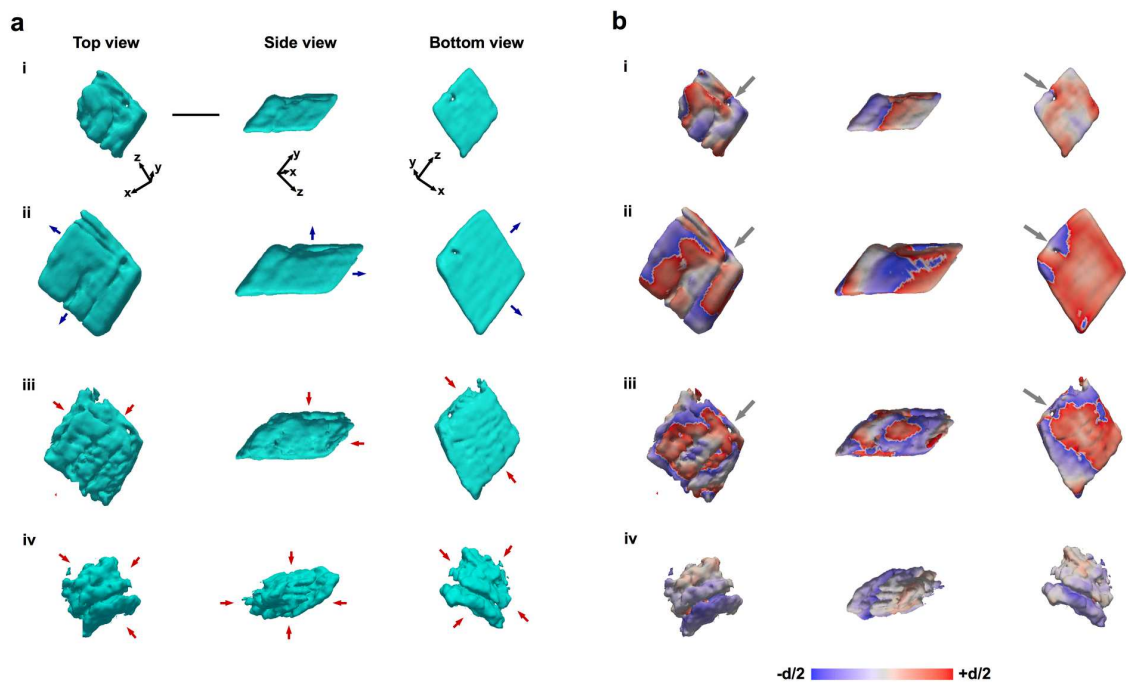}
 	\caption{\textbf{Growth and dissolution of calcite observed by BCDI} 
	Shown is an iso-surface rendering of the (a) electron density (reconstructed amplitude) and (b) projected displacement (phase) from (i) initially deposited calcite crystal, (ii) after secondary growth, and (iii $\&$ iv) after consecutive dissolution steps.  The scale bar is 1 $\mu$m.  Three different viewing angles of the crystal are shown in (a) and (b).  Given are top down (left), side (middle panel) and bottom up (right) perspectives of the imaged calcite crystal sitting flat on the substrate surface.  Perspectives highlight the shape transition that occurs during growth (i-ii) and dissolution (ii-iv).  Prominent surface advance (growth, blue arrows) and retreat (dissolution, red arrows) directions are shown.  The gray arrow points towards the primary dislocation continuously identifiable during crystal growth and dissolution (i-iii).  The beam direction is along the z axis, with the y axis vertical.  The sample/substrate is located at a set scattering angle towards the beam direction (z) and spanned plane (x-z). }\label{fig:1}
\end{figure}

\begin{figure}
 	\centering
	\includegraphics[width=.9  \textwidth]{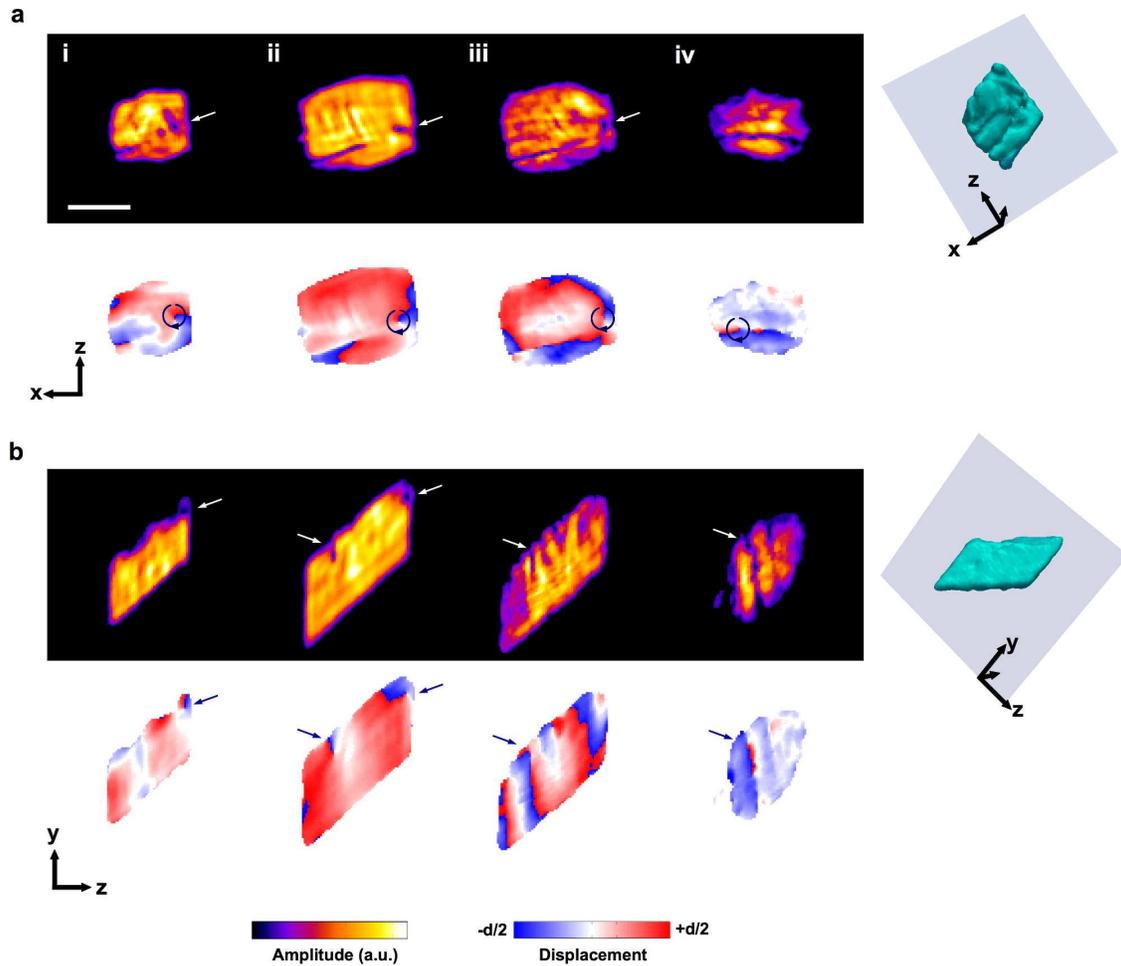}
 	\caption{\textbf{Slices showing the electron density and projected displacement during growth and dissolution} Shown is the initial crystal (i), after growth (ii) and repetitive dissolution steps (iii $\&$ iv) for two viewing directions (a) and (b).  The amplitude is shown in the top row (black background) with the dislocations highlighted by white arrows featuring a low-amplitude core.  The phase presented in the bottom row also shows the present, selected dislocations highlighted by the dark blue arrows.  Particularly evident is the spiral phase (displacement) that is characteristic of a screw dislocation.  The iso-surface to the right shows the location of the cut-planes.  The scale bar is 1 $\mu$m. }\label{fig:2}
\end{figure}

\begin{figure}[ht]
 	\centering
	\includegraphics[width=.95  \textwidth]{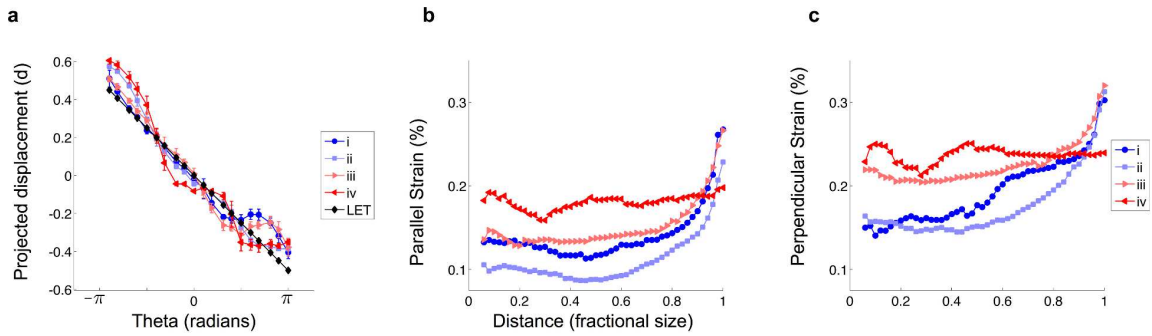}
 	\caption{\textbf{Displacement and strain line plots.  } Plotted in (a) is the recovered projected displacement measured at development stages (i-iv) as a function of theta for the dark blue circles with arrows from Fig. 2A.  This is compared to the displacement expected by linear elasticity theory (LET).  (b) The strain component parallel and perpendicular (c) to the scattering vector over the growth/dissolution cycle (i-iv) plotted as a function of fractional crystal size (center of the crystal 0 to crystal surfaces 1). This graph highlights the diminishing relevance of surface effects with (ii) growth of the crystal, and its increasing relevance with (iii $\&$ iv) dissolution.  The strain is calculated as the magnitude of the gradient of the displacement and averaged over successively larger shells.}\label{fig:3}
\end{figure}

\begin{figure}[h]
 	\centering
	\includegraphics[width=.9  \textwidth]{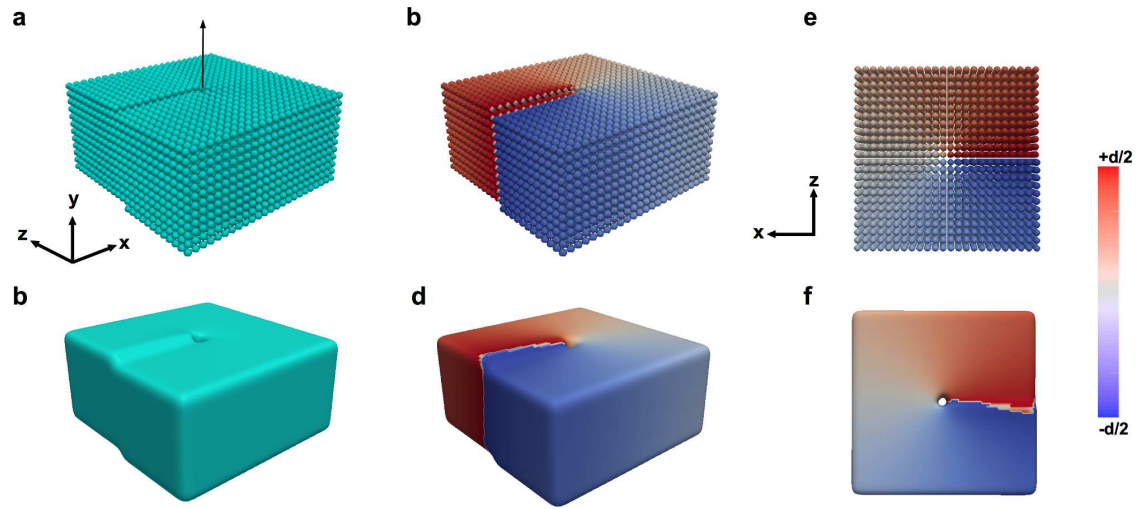}
 	\caption{\textbf{Simulation of a screw dislocation}  Iso-surface rendering of a screw dislocation (a) with atomic resolution, and by filtering (b) the Fourier transform of (a) with a Gaussian centred on the (0,1,0) Bragg peak, replicating the BCDI experiment.  The displacement is rendered onto the iso-surface for the atomic resolution (c) and the phase is rendered onto the surface for the BCDI simulation (d) showing the spiral phase centred around the dislocation core.  (e) and (f) show another view of the dislocation for the atomic resolution (e) and BCDI experiment (f), revealing the low density region at the core in the BCDI experiment simulation.  It should be noted that for the atomic resolution the displacement is mapped onto the iso-surface where for the BCDI example, the resultant phase is mapped on to the iso-surface.   }\label{fig:4}
\end{figure}

\begin{figure}[h]
 	\centering
	\includegraphics[width=.9 \textwidth]{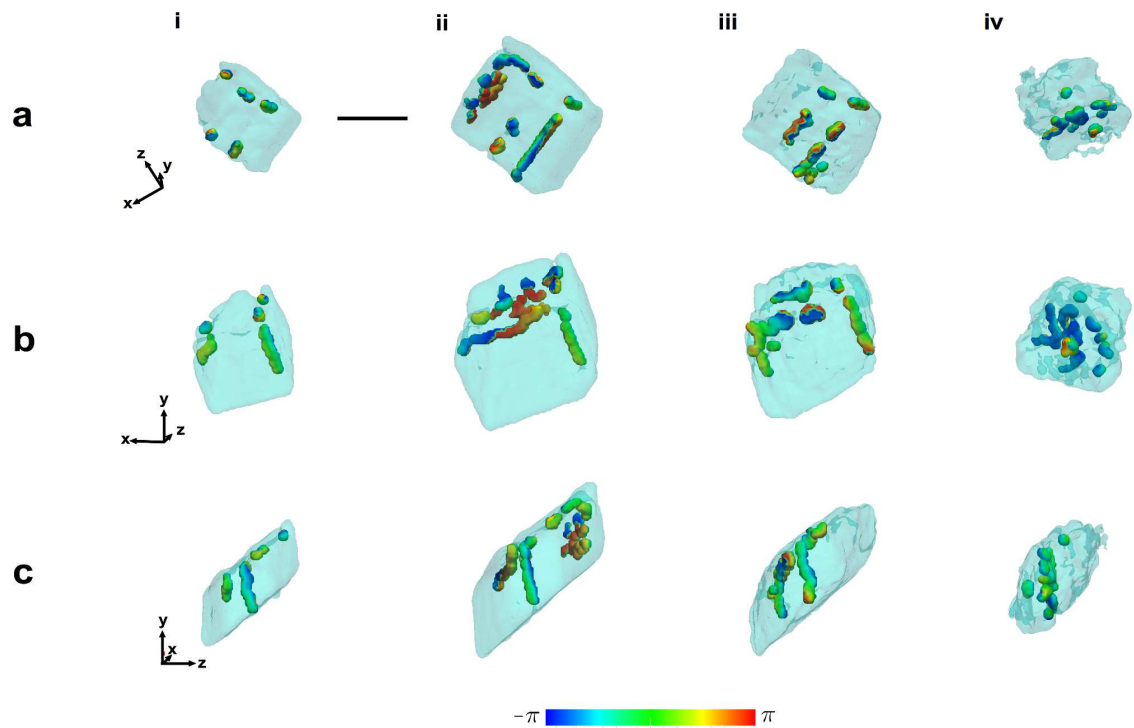}
\caption{\textbf{Iso-surface rendering of defect network within a calcite crystal.} Iso-surface renderings of dislocations present within the crystal imaged (i) before, and after (ii) growth / (iii-iv) dissolution.  The scale bar is 1 $\mu$m.  The evolution of the dislocations is evident through crystal growth (i to ii) and dissolution (ii-iii, iii-iv), where these are shown from left to right.  Three different viewing angles of the crystals are provided, (a) from base, (b) from top and (c) from side.  Dislocations are predominantly located near the surface and on the faces that grow most rapidly.  The phase has been mapped to the iso-surfaces of present dislocations, showing the characteristic spiral. }\label{fig:5}
\end{figure}

\clearpage

\begin{methods}

\subsection{Materials and General Preparative Methods}
Analytical grade, CaCl$_2$$\cdot$2H$_2$O, Acetic Acid, Urea and Urease (canavalia ensiformis, subunit molecular weight: $\sim$90.770 kDa) were purchased from Sigma-Aldrich and were used as received.  Aqueous solutions were prepared using Milli-Q Standard 18.2 M$\Omega$cm.  Solutions of 11-Mercapto-1-undecanol (Sigma-Aldrich) were freshly prepared in laboratory grade ethanol, and experiments were performed at a constant temperature of 21$^{\circ}\mathrm{C}$.  Reagent containing glassware was soaked overnight in 10$\%$ w/v NaOH, followed by rinsing with dilute HCl and washing with Milli-Q water.

\subsection{Substrate Preparation}
Glass slides and crystallizing dishes were placed overnight in Piranha solution (70:30 $\%$wt. sulfuric acid: hydrogen peroxide) and were then washed copiously with Milli-Q water before being exposed to the crystallisation solution.  Functionalized self-assembled monolayers (SAMs) were prepared on freshly deposited noble metal films.  Thin films were deposited either on silicon wafers using a Mantis Qprep 250 deposition system or a Cressington 308R coating system at a base pressure below 10$^{-6}$ mbar.  2 nm of Cr were initially deposited to promote substrate adhesion, followed by the evaporation of 50 nm of Au (Goodfellow, 99.99$\%$) at $\leq$0.1 nm/s.  SAMs were then prepared by immersing a prepared metal substrate in 1 mM thiol solution in ethanol, at room temperature for 24 h in the dark.  The SAMS were then thoroughly rinsed with ethanol and Milli-Q water and were dried under nitrogen before usage.

\subsection{Mineral Deposition}

Preferentially oriented calcite was obtained by adding 350 $\mu$l urease (1 mg mL$^{-1}$) to 1 mL of 5 mM CaCl$_2$ / 20 mM urea.  Separate droplets (100 $\mu$L) of the resulting aqueous solution were subsequently placed on the prepared SAM, and stored at 100$\%$ r.h..  Hydroxyl-terminated SAMs were used as the crystallization substrate to obtain calcite rhombohedra that were principally $\{$104$\}$ oriented.  Enzymatic hydrolysis of urea to ammonium and carbonate creates the required supersaturation profile, where the gradual increase in supersaturation ensures that a sufficient density of single crystal particles of CaCO$_3$ is obtained.  

\subsection{Overgrowth of Initial Deposits}

Observations of crystal growth were carried out on samples which had previously been centered into the Bragg condition.  A 50 $\mu l$ volume of $\approx$ 1 mM calcium bicarbonate solution was pipetted onto the sample, where evaporation and CO$_2$ out-gassing results in a supersaturation increase which causes crystal growth.  After evaporation of the droplet, the selected crystals were then re-analyzed.  
Calcium bicarbonate solutions were prepared according to the method of Kitano \cite{Kitano:1962aa}.  100 mg of CaCO$_3$ were added to one litre of Milli-Q water.  CO$_{2(g)}$ was then bubbled through this solution for three hours followed by filtering through a 200 nm Isopore GTTP membrane filter (Millipore) resulting in an initial pH of 6.4.  

\subsection{Partial Dissolution}
Dissolution of BCDI imaged calcite crystals was achieved by depositing 50 $\mu$L of 0.1 wt$\%$ acetic acid solution onto a 0.4 cm$^2$ substrate with an estimated number density of 0.1 crystals/ $\mu$m$^2$ with an average size of 1 $\mu$m.  The solution was then removed after 60 seconds.  This was followed by addition and removal of a drop of ethanol to wash the sample, and a further diffraction pattern was collected.  The process was then repeated between two and three times to obtain diffraction data for successive dissolution stages of the initially imaged single crystal.

\subsection{Enzymatic Hydrolysis (Urea-Urease) catalyzed Precipitation}
Calcium carbonate deposition using the enzymatic hydrolysis of urea by urease, and droplets placed on the substrate, allows controlled precipitation of crystals independent of droplet wetting behavior and gas-liquid interface area \cite{Ihli:2013aa}.  By variation of the urea and urease concentrations, which determines the rate of production of carbonate and ammonium, the rate of production of CaCO$_3$ can be controlled.  A slow increase in supersaturation was required to give crystals that were principally unstrained and which were present in a high enough number density to enable crystals to be easily located on the substrate during BCDI imaging.  The key underlying reactions are presented below.

\begin{align}
((NH_2)_2CO)+2H_2O & \xrightarrow{Urease} 2(NH_4^{+})+(CO_3^{2-}) \\
((NH_2)_2CO)+H_2O & \xrightarrow{or} 2(NH_3)+(CO_2) \\
(NH_3)+H_2O& \leftrightarrow (NH_4^{+})+{OH^{-}} \\
(CO_2^{-})+H_2O&\leftrightarrow(H_2CO_3) \\
(H_2CO_3)&\leftrightarrow(HCO_3^{-})+(H^{+})  \\
(HCO_3^{-})&\leftrightarrow(CO_3^{2-})+(H^{+}) \\
(CO_3^{2-})+({Ca^{2+}})&\leftrightarrow CaCO_3
\end{align}

\subsection{External Characterization}
CaCO3 precipitates at different growth and dissolution stages were characterised by Scanning Electron Microscopy (SEM).  Electron micrographs of uncoated specimen were obtained using a FEI Nova NanoSEM 650.  Crystal growth and dissolution rates of outgrown individual calcite crystal were obtained using an inverted Olympus IX-70 confocal microscope.

\subsection{Experimental Setup}

BCDI experiments were performed at Beamline I-16 at the Diamond Light Source (DLS) UK.  An undulator produced X-rays which were monochromatized to an energy of 8 keV using a channel-cut Si(111) monochromator.  Calcite crystals supported on substrates were placed on a diffractometer which had its rotation centre aligned with the X-ray beam.  Slits were used to reduce the area illuminated by the X-rays.  An X-ray sensitive, CMOS pixel detector (Medipix3 ) with 256$\times$256 square pixels of side length 55 $\mu$m was positioned at the desired diffraction angle for an off-specular $\{$104$\}$ reflection at a distance of 2.7 m from the sample. An evacuated flight tube was placed between the detector and sample to reduce air absorption and scatter.  To measure its full 3D diffraction pattern, the crystal was rotated by 0.3 degree with a 0.003 degree step size.  A two-dimensional slice of the 3D far-field diffraction pattern was recorded at each angle of rotation.  By stacking all of these two-dimensional diffraction frames together, a complete 3D diffraction pattern was obtained, from which real-space images can be reconstructed (see Phase Retreival).  Due to the small size of the crystals, $\sim$ 1-2 $\mu$m, the illumination can be considered to be almost completely coherent.  To ensure that the same crystal could be analysed over a growth/ dissolution cycle, a local search in reciprocal space was performed to pick an isolated single crystal.  This ensured that there was little chance of a signal originating from another crystal overlapping the original crystal diffraction.

\subsection{Data preparation}
For each growth/dissolution stage, the crystal was rotated through its rocking curve where a series of 2D diffraction patterns were collected.  A total of 121 patterns were collected along the rocking curve with a separation of 0.003 degrees.  The exposure time for each pattern in the rocking curve was 1 second.  This process (collecting 2D diffraction along the rocking curve) was repeated three times for each growth/dissolution stage.  This resulted in three data sets for each stage which were then summed together to form a single 3D diffraction pattern which can be seen in Supplementary Information Fig. 1.  The alignment of the three sets was done by calculating their centre of masses and shifting the data sets to the centre of mass of the first data set (to the nearest pixel).  This data set was then thresholded, where all pixels with 2 or less photons were set to 0.  Several pixels on the detector were reading high values even when no photons were present (`hot pixels').  These pixels had their values set to 0.  In each of the two detector pixel directions, the data were binned by a factor of 2 and cropped to a final size of 80 pixels while in the scan direction the data was not binned and was padded (with zeros) to a final size of 128 pixels.  
After the initial growth stage, diffraction from another (different) crystal was observed on the detector.  This diffraction was identified as coming from a different crystallite by the fact it was spatially separated on the detector (from the original diffraction) with no visible interference fringes between this diffraction and the original.  This diffraction was removed by setting it to 0. 
This summed, thresholded, zeroed and binned/padded data set was then used for the reconstruction.

\end{methods}


\newpage

\end{document}